\documentclass{report}
\usepackage[utf8]{inputenc}
\usepackage{graphicx}
\usepackage{rotfloat}
\usepackage{rotating}
\usepackage{booktabs}
\usepackage{stfloats}
\usepackage{tabu}
\usepackage{subcaption}
\usepackage{amsmath}
\usepackage{comment}
\usepackage[a4paper, total={6in, 8in}]{geometry}
\usepackage{setspace}
\usepackage{pdfpages}
\usepackage{mathtools}
\newtagform{brackets}{[}{]}
\usetagform{brackets}

\title{Supporting Information}

\date{}

\begin{document}


# Modelling an equivalent b-value in diffusion-weighted steady-state free precession

Benjamin C. Tendler[1], Sean Foxley[2], Michiel Cottaar[1], Saad Jbabdi[1,†], Karla Miller[1,†]

[1]Wellcome Centre for Integrative Neuroimaging, FMRIB, Nuffield Department of Clinical Neurosciences, University of Oxford, Oxford, United Kingdom

[2]Department of Radiology, University of Chicago, Chicago, IL, United States

†Equal contribution

**Abstract**

Purpose: Diffusion-weighted steady-state free precession (DW-SSFP) is shown to provide a means to probe non-Gaussian diffusion through manipulation of the flip angle. A framework is presented to define an effective b-value in DW-SSFP.

Theory: The DW-SSFP signal is a summation of coherence pathways with different b-values. The relative contribution of each pathway is dictated by the flip angle. This leads to an apparent diffusion coefficient (ADC) estimate that depends on the flip angle in non-Gaussian diffusion regimes. By acquiring DW-SSFP data at multiple flip angles and modelling the variation in ADC for a given form of non-Gaussianity, the ADC can be estimated at a well-defined effective b-value.

Methods: A gamma distribution is used to model non-Gaussian diffusion, embedded in the Buxton signal model for DW-SSFP. Monte-Carlo simulations of non-Gaussian diffusion in DW-SSFP and diffusion-weighted spin-echo (DW-SE) sequences are used to verify the proposed framework. Dependence of ADC on flip angle in DW-SSFP is verified with experimental measurements in a whole, human post-mortem brain.

Results: Monte-Carlo simulations reveal excellent agreement between ADCs estimated with DW-SE and the proposed framework. Experimental ADC estimates vary as a function of flip angle over the corpus callosum of the postmortem brain, estimating the mean and standard deviation of the gamma distribution as $1.50 \cdot 10^{-4}$ mm$^2$/s and $2.10 \cdot 10^{-4}$ mm$^2$/s.

Conclusion: DW-SSFP can be used to investigate non-Gaussian diffusion by varying the flip angle. By fitting a model of non-Gaussian diffusion, the ADC in DW-SSFP can be estimated at an effective b-value, comparable to more conventional diffusion sequences.

Keywords: diffusion-weighted steady-state free precession, diffusion-weighted spin-echo, b-value, non-Gaussian diffusion, Monte-Carlo, postmortem MRI

**Introduction**

Diffusion-weighted steady-state free precession (DW-SSFP) is a powerful sequence that achieves strong diffusion weighting by maintaining a steady-state in which magnetisation accumulates diffusion contrast over multiple TRs (1–4). The DW-SSFP sequence for each TR consists of a single RF pulse and single diffusion gradient followed by signal acquisition (Fig. 1a; this depiction neglects imaging gradients, which are generally refocused to have zero net area). The DW-SSFP sequence has many favourable properties for probing the diffusion properties of tissue (5,6): it is very SNR-efficient (1,7), generates strong diffusion weighting in MR systems with limited gradient strengths (1,8,9) and yields high-SNR diffusivity estimates in samples with short $T_2$ (7,10,11). These properties stem from the steady-state nature of the sequence (5). In DW-SSFP, transverse magnetization is not spoiled between RF pulses and the short TR (typically TR<$T_2$) ensures transverse magnetisation persists over multiple excitations, leading to numerous signal-forming coherence pathways (12,13). The signal received from coherence pathways with high b-values (8,14) leads to strong diffusion weighting. The short TR prevents relaxation from destroying the available signal before sampling. Although this saturates the magnetisation, the large fraction of each TR spent acquiring signal provides a high-SNR efficiency.

The DW-SSFP sequence has two major challenges to overcome (5): first, it is very sensitive to motion; second, it does not have a well-defined b-value. One environment where the properties of DW-SSFP are very well suited is imaging of fixed, post-mortem tissue, which is devoid of motion but plagued by low $T_2$ and reduced diffusion coefficients (7,15). Nevertheless, interpretation of these post-mortem data suffer from the lack of a well-defined b-value, which is a direct result of the signal reflecting a summation of numerous coherence pathways, each with a different b-value (5,8). As diffusive motion in tissue is generally non-Gaussian, this poorly-defined b-value prevents comparisons between diffusivity estimates obtained with the DW-SSFP and more conventional measurements using the diffusion-weighted spin-echo (DW-SE) sequence (Fig. 1b).

Formation of the steady state in DW-SSFP is a function of both experimental parameters (flip angle and TR) and sample properties (relaxation and diffusivity). However, unlike the DW-SE sequence, the diffusion weighted terms in DW-SSFP are not readily separable as a simple multiplicative term (8). Instead, alterations in the prescribed flip angle, TR, and relaxation times alter the relative weighting of each coherence pathway, and hence result in a different diffusion weighting (5). This surprising result highlights the fact that in

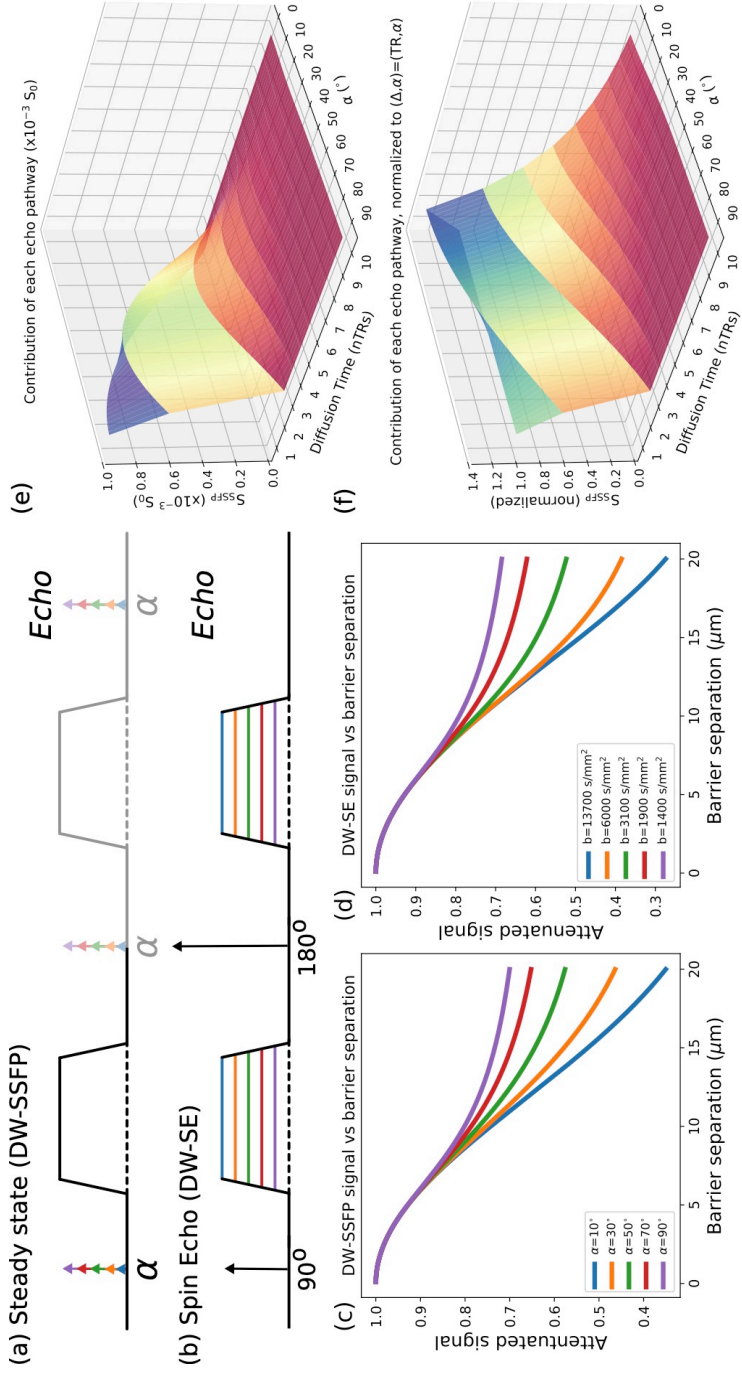

Figure 1: Comparison of DW-SSFP and DW-SE sequences under non-Gaussian diffusion (e.g. due to restricting barriers). (a) The DW-SSFP sequence consists of a single RF pulse and diffusion gradient per TR that achieves the equivalent of Stejskal-Tanner gradient pairs over multiple TR periods. (b) In systems with non-Gaussian diffusion, here represented by diffusion restricted between two parallel barriers, changing the flip angle results in a different level of signal attenuation. (b) With the DW-SE sequence, gradient pairs are explicitly included on either side of the refocusing pulse. (d) By changing the strength of the applied diffusion gradient, we can achieve a similar sensitivity to restricted diffusion. (e) DW-SSFP signal can be represented as a summation of coherence pathways, each of which has a well-defined diffusion time representing the number of TR periods between the experienced diffusion gradients. The relative contribution of coherence pathways is a function of flip angle. (f) Normalising these coherence pathways in (e) to the pathway with $\Delta$ = TR (i.e. the spin-echo pathway) provides insight into the sensitivity of DW-SSFP to non-Gaussian diffusion. Reducing the flip angle increases the contribution of coherence pathways with longer diffusion times, which are more sensitive to non-Gaussian diffusion. All simulations performed under the two transverse period approximation (Eq. [1]), defining $D = 3.5 \cdot 10^{-4}$ mm$^2$/s, TR = 28.2 ms, $T_1$ = 600 ms and $T_2$ = 20 ms. (c) and (d) are simulated with a model of diffusion restricted between two parallel barriers as described in (16). (e) and (f) are calculated assuming Gaussian diffusion.

3

DW-SSFP, there is not a standalone diffusion preparation (gradients and their timings) that determines the degree of diffusion weighting, as is the case for spin and stimulated echo sequences. Looking at this from a different perspective, the idiosyncrasies of the DW-SSFP signal formation mechanism present us with an opportunity: to probe the diffusion properties of tissue without any modification to the diffusion encoding gradients. Figures 1c and d simulate the received DW-SSFP and DW-SE signal for diffusion restricted between two parallel barriers. Signal attenuation is altered by changing the flip angle in DW-SSFP (Fig. 1c), similar to changing the b-value in DW-SE (Fig. 1d).

In this work, we show that we can probe different diffusion time (and therefore b-value) regimes by varying the flip-angle in DW-SSFP. As with varying b-values in more conventional diffusion measurements, this flip angle dependence changes the apparent diffusion coefficient (ADC) estimates in systems with non-Gaussian diffusion. Based on this concept, we propose a method to translate quantitative diffusivity estimates derived with DW-SSFP, in which b-values are not well defined, into ADC estimates at a single effective b-value, as would be measured using more conventional sequences such as DW-SE. This is achieved by defining DW-SSFP signal behaviour under a model of non-Gaussianity and translating the measured DW-SSFP signal at multiple flip angles into an ADC at an equivalent, well-defined b-value. The specific model presented here combines a gamma variate distribution of diffusivities with the Buxton model of DW-SSFP signal (8), but this approach can be adapted to other forms of non-Gaussianity (17) and alternative signal models (4,18). The derived signal model is verified with Monte-Carlo simulations of both DW-SSFP and DW-SE signal evolution, and the expected signal dependence is demonstrated using DW-SSFP datasets acquired at multiple flip angles in post-mortem brain tissue.

**Theory**

*The two transverse-period approximation*

The two transverse-period approximation of DW-SSFP (6,8) is signal model that makes the simplifying assumption that coherence pathways do not survive beyond two periods in the transverse plane. This approximation, considered valid when TR $\geq$ ~1.5·$T_2$ (8), is particularly helpful for building intuition into the dependence of diffusion times on flip angle. Under these conditions, the DW-SSFP signal can be described as the weighted sum of spin- and stimulated-echo pathways:

$$S_{\text{SSFP}}(\alpha, T_1, T_2, \text{TR}, q, D) =$$

$$\frac{-S_0(1-E_1)E_1 E_2^2 \sin\alpha}{2(1-E_1\cos\alpha)} \cdot \left[ \underbrace{\frac{1-\cos\alpha}{E_1} A_1}_{\text{Spin echo}} + \sin^2\alpha \underbrace{\sum_{n=1}^{\infty}(E_1\cos\alpha)^{n-1} A_1^{n+1}}_{\text{Stimulated echoes}} \right], \quad [1]$$

where $S_0$ is the equilibrium magnetization, $E_1 = e^{-\frac{\text{TR}}{T_1}}$, $E_2 = e^{-\frac{\text{TR}}{T_2}}$, $\alpha$ is the flip angle, $A_1 = e^{-q^2 \cdot \text{TR} \cdot D}$, $D$ is the diffusion coefficient and $q = \gamma G \tau$, where $\gamma$ is the gyromagnetic ratio, $G$ is the diffusion gradient amplitude and $\tau$ is the diffusion gradient duration. In Eq. [1], the first term in the square brackets represents a spin-echo pathway (i.e. the magnetisation that is in the transverse plane in two consecutive TRs), and the second term describes the stimulated-echo pathways, each characterised by two transverse periods separated by *n* longitudinal periods. The diffusion time, $\Delta$, is well defined for each individual pathway (spin echo: $\Delta = \text{TR}$, stimulated echo: $\Delta = (n+1) \cdot \text{TR}$). The effect of diffusion time is embodied in the $A_1$ terms, with each pathway attenuated by $e^{-q^2 \cdot \Delta \cdot D}$. Under the two transverse-period approximation, the signal is a weighted sum of contributions from different pathways with different diffusion times, with relative signal weights that depend on the flip angle ($\alpha$), TR and $T_1$. Changes in $T_2$ do not alter the relative weighting of each pathway, since the assumption is that only coherence pathways with two transverse periods contribute signal. Example pathways considered in the two-transverse approximation are illustrated in Supporting Information Figs. S1a-c.

Figure 1e visualises the signal contributions of each pathway (amplitudes calculated from individual terms in the summation in Eq. [1]). Pathways with longer diffusion times lead to signals that are more diffusion weighted and informative about restrictive diffusion. At intermediate flip angles the overall signal contribution from the different pathways peaks, leading to increased SNR. We can visualise the relative contributions of different pathways at a given flip angle by normalising to the signal from the spin-echo ($\Delta = 1 \cdot \text{TR}$) pathway (Fig. 1f). This normalisation makes it clear that decreasing the flip angle increases the relative contribution of simulated-echo pathways with longer diffusion times, leading to an increase in diffusion contrast. However, this comes at a tradeoff of overall signal levels as shown in Fig. 1e, leading to a peak in contrast-to-noise ratio at an intermediate flip angle (7).

The two transverse-period approximation provides an intuitive way to see that changing the flip angle in DW-SSFP alters the diffusion time regime that the signal is sensitive to, with an increased flip angle corresponding to a shorter effective diffusion time.

The DW-SSFP signal can be thought as a temporally-blurred mixture of the "cleaner" diffusion time behaviour that is captured by more conventional DW-SE (or DW-STE) signals, corresponding to a single point on the $\Delta$ axis.

*The full Buxton model of DW-SSFP*

The full Buxton model of DW-SSFP (6,8) accounts for all coherence pathways, including those that survive more than 2 TRs in the transverse plane. Summing over all coherence pathways yields the expression:

$$S_{SSFP}(\alpha, T_1, T_2, TR, q, D) = -\frac{S_0(1-E_1)E_2 A_2^{-\frac{2}{3}}\left(F_1 - E_2 A_1 A_2^{\frac{2}{3}}\right)\sin\alpha}{r - F_1 s}, \qquad [2]$$

where definitions of terms $r$, $s$ and $F_1$ are provided in the Appendix and $A_2 = e^{-q^2 \cdot \tau \cdot D}$. This more complete model allows for the existence of additional coherence pathways, including pathways that remain in the transverse plane over multiple TRs, and coherence pathways that give rise to multiple signal forming echoes over their lifetime (14). Unlike the two transverse approximation, the Buxton model accounts for pathways experiencing more than two diffusion gradients, including some with an effective q-value that is an even multiple of the q in a single TR period. Additionally, the relative signal weighting of pathways is dependent on $T_2$, unlike the two-transverse approximation (14). Examples of these additional pathways are given in Supporting Information Fig. S1. Under the full Buxton model, we therefore lose a strict correspondence between pathway and diffusion time; instead, changing the flip angle is equivalent to probing different b-value regimes, with smaller effective b-value at higher flip angle. The DW-SSFP signal is a blurred mixture of the "cleaner" b-value behaviour that is captured by more conventional DW-SE (or DW-STE) signals.

*Investigating Non-Gaussianity*

Diffusion in tissue is restricted and hindered by membranes, causing the ADC at higher b-values to be less than one would predict using the Gaussian propagator describing free diffusion. As can be inferred from Figs. 1c-f, any dependence of diffusivity on b-value will give rise to variable apparent diffusion coefficients (ADCs) for different flip angles in DW-SSFP. Hence, while conventional sequences typically characterise non-Gaussian diffusion using measurements at multiple diffusion times or q-values, this can also be accomplished in DW-SSFP through measurements at multiple flip angles. This also provides a route to

address the poorly-defined b-value in a DW-SSFP measurement through translation into a more conventional framework with a well-defined b-value.

We demonstrate this concept using a gamma distribution of diffusivities (Fig. 2a) to describe non-Gaussian diffusion (19,20). The gamma distribution, $\rho(D; D_m, D_s)$, can be described in terms of a mean, $D_m$, and a standard-deviation, $D_s$. For DW-SE, the signal for a gamma distribution of diffusivities is defined as (19,20):

$$S_{\text{SE},\Gamma}(b, D_m, D_s) = S_0 \int_0^\infty e^{-bD} \rho(D; D_m, D_s) \, dD \quad [3]$$

$$= S_0 \left( \frac{D_m}{D_m + bD_s^2} \right)^{\frac{D_m^2}{D_s^2}},$$

where $b = q^2 \cdot (\Delta - \tau/3)$. This distribution of diffusivities can be embedded in the full Buxton signal model of DW-SSFP as:

$$S_{\text{SSFP},\Gamma}(\alpha, T_1, T_2, \text{TR}, q, D_m, D_s) =$$
$$-S_0(1 - E_1)E_2 \sin\alpha \int_0^\infty \frac{A_2^{-2/3}(F_1 - E_2 A_1 A_2^{2/3})}{r - F_1 s} \rho(D; D_m, D_s) \, dD. \quad [4]$$

The integral describing DW-SSFP can be evaluated using numerical integration. Figure 2b depicts how the ADC varies as a function of flip angle for the three different gamma distributions in Fig. 2a. As we increase the flip angle, we obtain a higher estimate of ADC, consistent with our expectations of an increased ADC estimate as we decrease the b-value.

*A framework to translate between DW-SSFP and DW-SE measurements*

Given quantification of the ADC in DW-SSFP, we can define an 'effective' b-value to be that which yields the same ADC estimate using the DW-SE sequence. Translating ADC estimates from DW-SSFP into an equivalent ADC at a single b-value can be achieved in the context of a common, underlying non-Gaussianity.

From diffusion-weighted DW-SSFP data obtained at multiple flip angles, the ADC can be uniquely determined at each flip angle by solving Eq. [2] (Fig. 2b - dots), given knowledge of the experimental protocol, $T_1$ and $T_2$. Our diffusion model (Fig. 2a) can be subsequently fitted to the multi-flip data (Eqs. [2] and [4]) to uniquely determine a value of $D_m$ and $D_s$ that can describe the evolution of ADC with flip angle (Fig. 2b – dashed lines). We can use the values of $D_m$ and $D_s$ to subsequently simulate the ADC at any given DW-SE b-value (Fig. 2c) by comparing Eq. [3] with the DW-SE signal under the Stejskal-Tanner model ($S = S_0 \exp(-bD)$). Alternatively, we can determine the equivalent b-value that

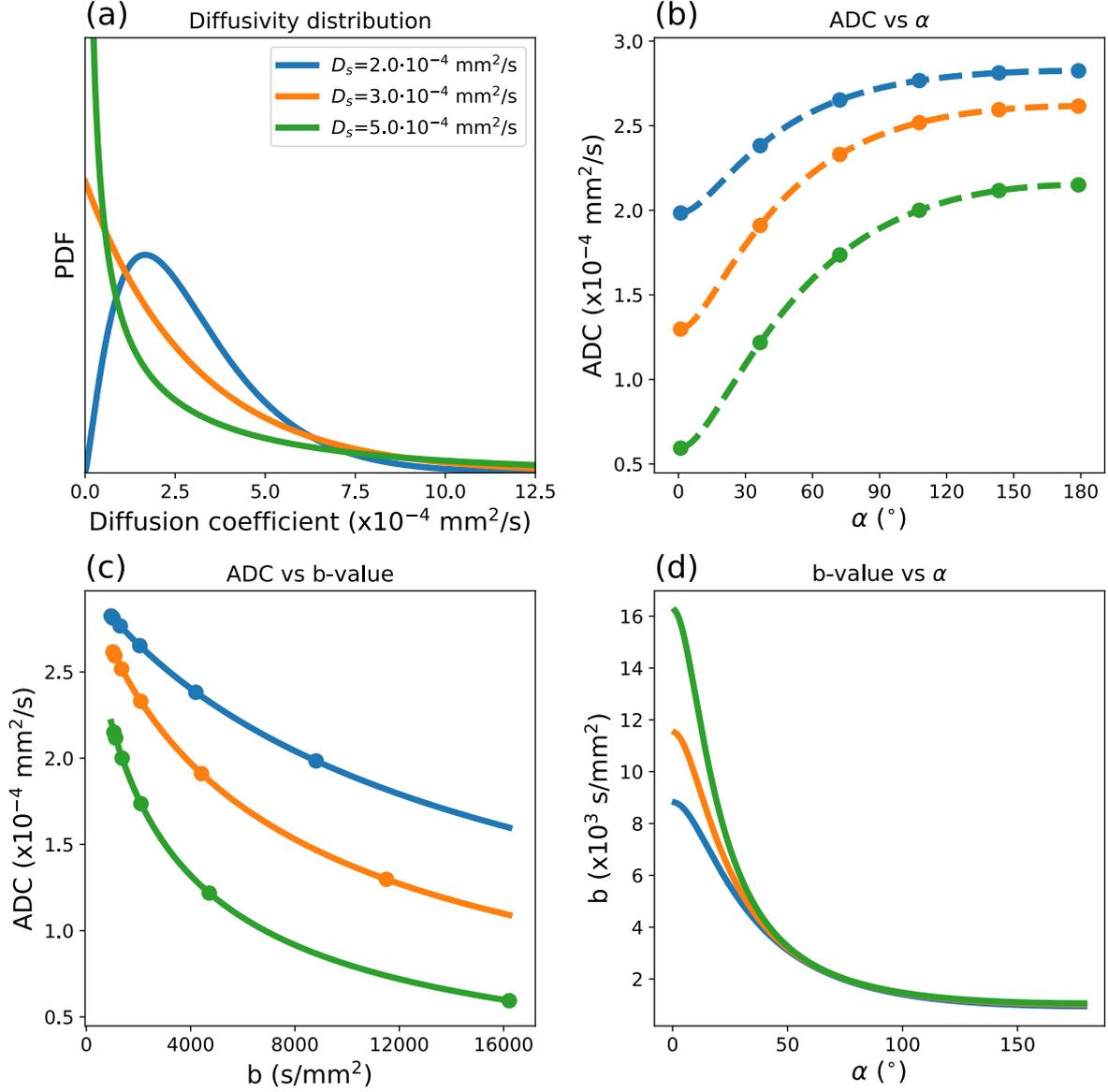

Figure 2: Three different gamma distributions (a) with $D_m = 3.0 \cdot 10^{-4}$ mm$^2$/s and $D_s$ defined as per the legend. (b) The associated evolution of ADC with flip angle for DW-SSFP under these distributions. The Buxton model for Gaussian diffusion can be fit to DW-SSFP measurements (b – dots) to obtain ADC estimates at multiple flip angles. Comparing Eqs. [2] and [4], we can then fit a gamma distribution to these ADC estimates (b – dashed lines). If we wish to translate this gamma variate to the equivalent ADC estimates that would be obtained from DW-SE, we can subsequently calculate the ADC for a given b-value assuming the same gamma distribution (c – solid lines). Alternatively, we can define a DW-SE b-value at any given DW-SSFP flip angle that gives rise to an equivalent ADC (d). Combining these expressions, we can plot the ADC estimates measured with DW-SSFP (b - dots) vs the DW-SE b-value (c - dots). Simulation performed over the range $\alpha = 1°$ - $179°$, setting the diffusion gradient amplitude = 5.2 G/cm & diffusion gradient duration = 13.56 ms (q=300 cm$^{-1}$), TR = 28.2 ms, $T_1$ = 600 ms and $T_2$ = 20 ms.

would yield the same estimate of ADC as measured with DW-SSFP (Fig. 2d) at a given flip angle. A detailed processing pipeline is provided in Supporting Information Fig. S2.

**Methods**

*Monte-Carlo simulations of DW-SSFP and DW-SE signal*

Uniformly distributed spin trajectories were generated using the Camino toolbox (21) ($5 \cdot 10^5$ spins, $D = 3.5 \cdot 10^{-4}$ mm$^2$/s, 250 time steps), modified to produce trajectories that followed a Gaussian distribution of displacements per time step rather than a step of fixed length (21). A gamma distribution of diffusivities was subsequently generated from the trajectories in MATLAB (version 2017a, The MathWorks, Inc., Natick, MA), setting $D_m = 1.50 \cdot 10^{-4}$ mm$^2$/s and $D_s = 2.10 \cdot 10^{-4}$ mm$^2$/s, consistent with the corpus callosum of the postmortem sample used in our experimental analysis (see following section and Results).

The DW-SSFP signal was simulated using in-house code written in MATLAB, with approximately the same parameters as used in our experimental measurements (see following section) (TR = 28.2 ms, diffusion gradient duration = 13.56 ms, diffusion gradient amplitude = 52 mT/m, q = 300 cm$^{-1}$, flip angles = 10º to 170º in 10º increments), setting $T_1 = 568$ ms and $T_2 = 19.8$ ms, the mean over the corpus callosum of our sample. Non-diffusion weighted DW-SSFP data was additionally simulated, setting $D = 0$ mm$^2$/s. A single time step corresponded to one TR of the DW-SSFP sequence.

A DW-SE signal was additionally simulated (diffusion gradient duration = 13.56 ms, Δ = 40 ms, b-values = 0 to 14000 s/mm$^2$ at 1000 s/mm$^2$ increments, achieved by increasing the gradient amplitude). A single time step corresponds to 0.4 ms.

*Experimental demonstration of the DW-SSFP flip angle dependency*

A whole postmortem brain sample was scanned on a 7T Siemens MR system (1Tx/32Rx head coil) with a DW-SSFP sequence for a single diffusion direction at multiple flip angles (resolution = 0.85 x 0.85 x 0.85 mm$^3$, TR = 28.2 ms, TE = 21 ms, BW = 393 Hz/pixel, diffusion gradient duration = 13.56 ms, diffusion gradient amplitude = 52 mT/m, $q = 300$ cm$^{-1}$, direction = [0.577,0.577,0.577], flip angles = 10º to 90º at 5º increments). At each flip angle, an equivalent non-diffusion weighted DW-SSFP dataset was acquired with a small diffusion gradient ($q = 20$ cm$^{-1}$) to ensure dephasing of the magnetisation and to prevent banding artefacts (22). $T_1$, $T_2$ and $B_1$ maps (23) were additionally acquired over the postmortem brain sample, which are required for accurate modelling of the DW-SSFP signal

(11). Details of these acquisitions and processing are provided in Supporting Information Table S1.

Prior to analysis, a Gibbs ringing correction was applied to the DW-SSFP images (24). To reduce noise floor bias, the mean background signal was estimated and removed from the DW-SSFP signal (25). All coregistrations within and between the DW-SSFP datasets and other imaging modalities were performed using a 6 degrees of freedom transformation with FSL FLIRT (26,27).

The voxelwise ADC was estimated over the corpus callosum at each nominal flip angle using Eq. [2]. To avoid fitting for $S_0$, the experimental diffusion weighted DW-SSFP data was normalised by the non-diffusion weighted DW-SSFP data and fit with $S_{\text{SSFP}}(\alpha, T_1, T_2, \text{TR}, q, \text{ADC})/S_{\text{SSFP}}(\alpha, T_1, T_2, \text{TR}, 0, \text{ADC})$ (noting $S_{\text{SSFP}}(\alpha, T_1, T_2, \text{TR}, 0, \text{ADC}) \neq S_0$). The mean ADC over the corpus callosum was subsequently calculated at each flip angle and fit to Eq. [4] to determine $D_m$ and $D_s$. Fitting was performed in Python (28) using the SciPy *curve_fit* function, implemented with the Levenberg-Marquardt algorithm (29). Numerical integration of Eq. [4] was performed using the SciPy *quad* command.

**Results**

*Monte-Carlo simulations of the DW-SSFP and DW-SE sequence*

Figures 3a and b compare the simulated signal attenuation of the DW-SSFP and DW-SE signal estimated for a gamma-variate distribution (blue circles) to forward calculations from Eqs. [3] and [4] (green lines). Fitting to the Monte-Carlo signals, we estimated $D_m = 1.48 \cdot 10^{-4}$ mm²/s and $D_s = 2.04 \cdot 10^{-4}$ mm²/s for DW-SSFP, and $D_m = 1.49 \cdot 10^{-4}$ mm²/s and $D_s = 2.10 \cdot 10^{-4}$ mm²/s for DW-SE, compared to the original values of $D_m = 1.50 \cdot 10^{-4}$ mm²/s and $D_s = 2.10 \cdot 10^{-4}$ mm²/s. Similar to acquiring DW-SE data at multiple b-values, these simulations suggest that the DW-SSFP signal acquired at multiple flip angles is able to encode non-Gaussian diffusion. Fitting a Gaussian model assuming a single diffusion coefficient (red line) is unable to provide an accurate fit to the simulated signals.

By calculating ADC estimates from the signal attenuation using the original Buxton model for DW-SSFP and the Stejskal-Tanner model for DW-SE (i.e. both assuming purely Gaussian diffusion, shown in Fig. 3c and d), we can determine the equivalent DW-SE b-value that corresponds to the ADC estimate at each DW-SSFP flip angle (Fig. 3e). These results highlight the substantial range of effective b-values achievable with the DW-SSFP sequence

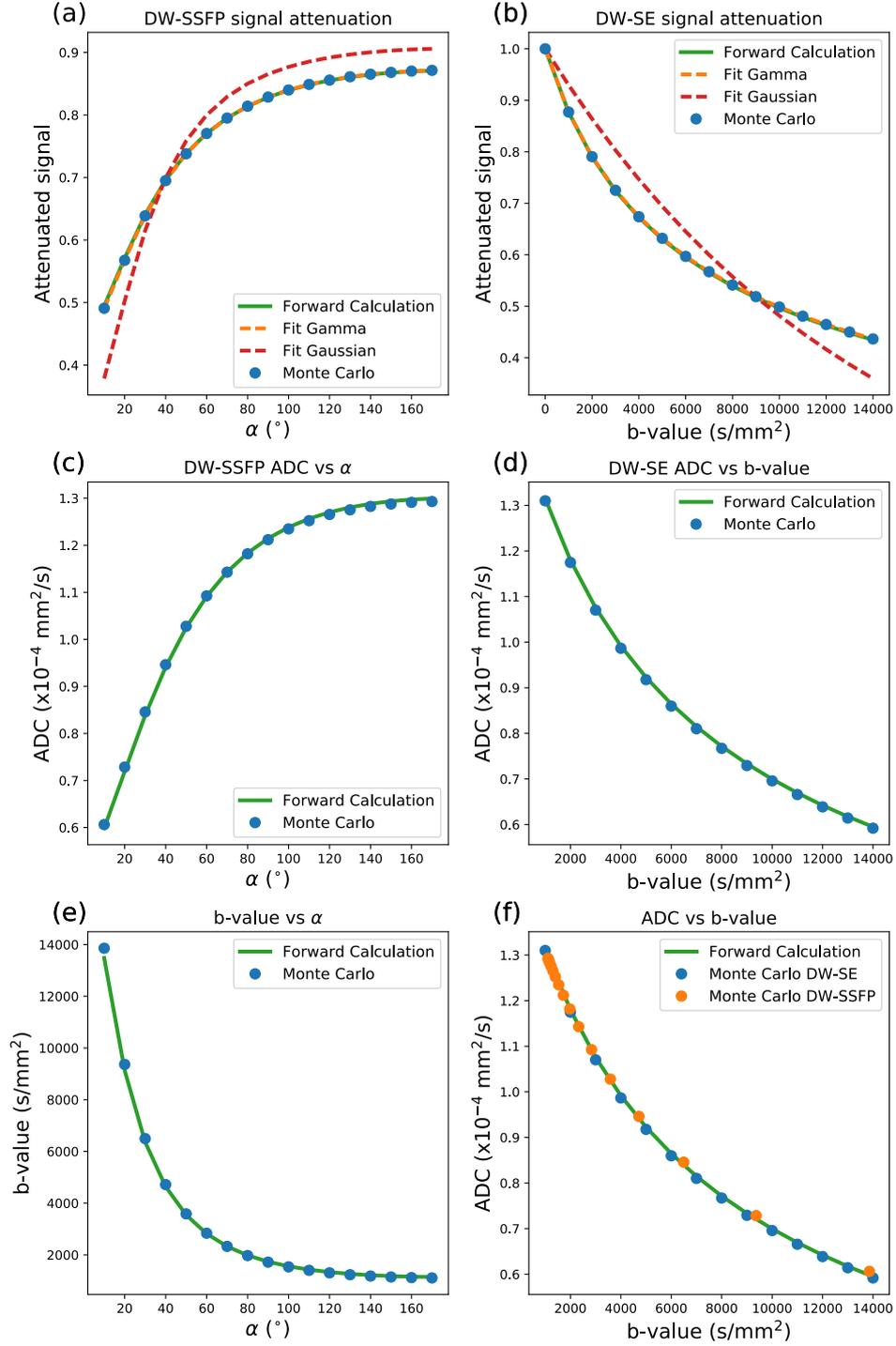

Figure 3: Results of the Monte-Carlo simulations. (a) and (b) reveals the signal attenuation for the DW-SSFP (a) and DW-SE (b) simulations respectively: Monte-Carlo simulations (blue dots), forward calculations of the DW-SSFP and DW-SE signal under a gamma-variate distribution (green lines), fits to the Monte-Carlo solutions (dashed orange lines) and fits assuming only a single diffusion coefficient (dashed red lines). (c) and (d) reveal how the estimated ADC (Eq. [2]) varies with DW-SSFP flip angle and DW-SE b-value. By comparing the ADC estimates in (c) and (d), we can determine which DW-SSFP flip angle gives rise to an equivalent ADC estimate (e). This allows us to transform our Monte-Carlo estimates of ADC with the DW-SSFP sequence into the same space as the DW-SE sequence (f).

by modifying the flip angle alone. With this, we are able to translate our DW-SSFP signal, which reflects a blurring of different signals with well-defined b-values, into a DW-SSFP ADC at a well-defined effective b-value, demonstrating the same ADC evolution as DW-SE data (Fig. 3f). Note that in Fig. 3f, the orange and blue data points were derived from separate simulations.

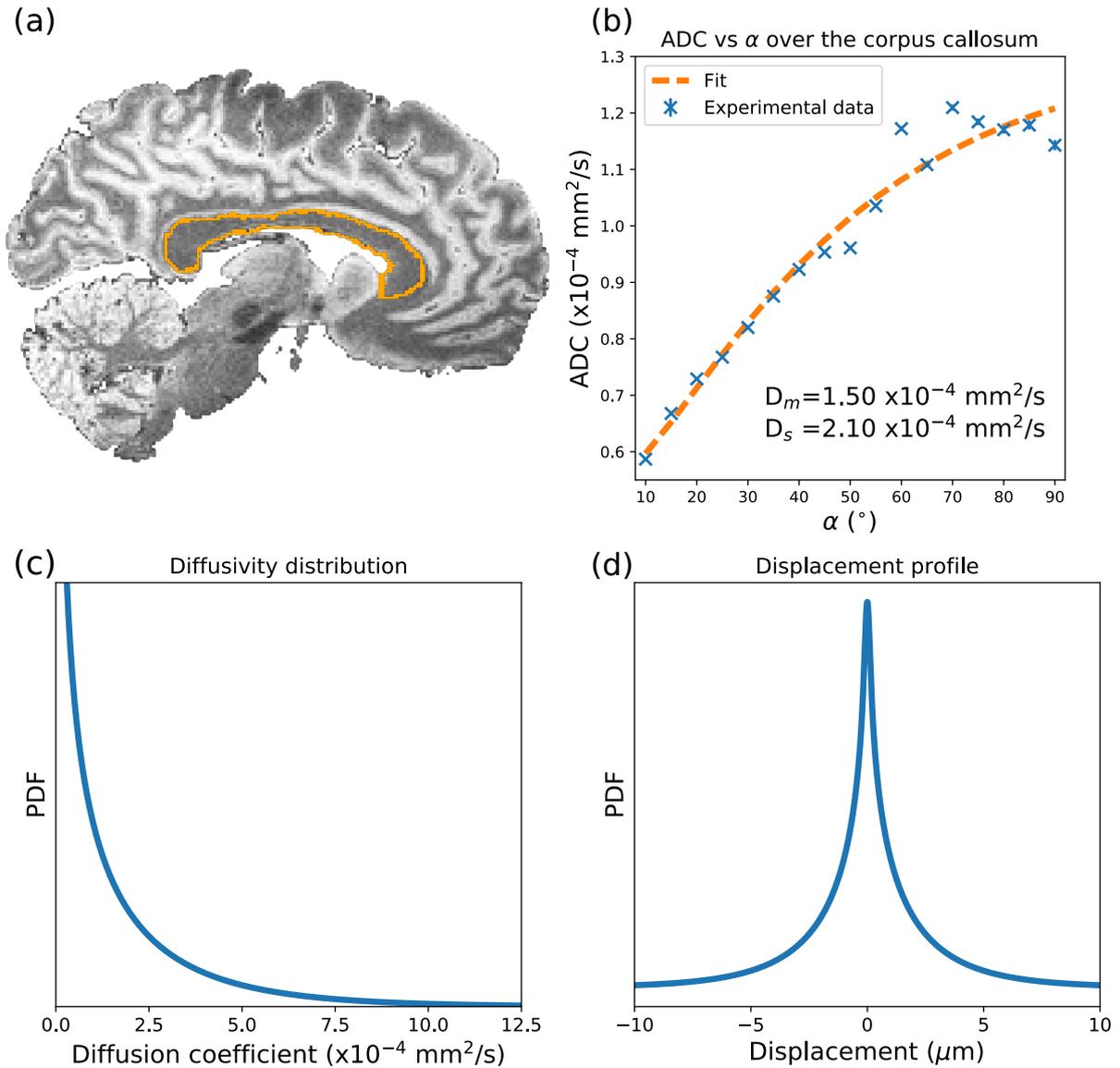

Figure 4: Sagittal slice of the DW-SSFP data (q=20cm$^{-1}$) acquired over the postmortem brain sample (a) with the corpus callosum outlined in orange. By determining the ADC at each flip angle (b – blue crosses) using Eq. [2], we demonstrate that the flip angle in DW-SSFP sensitises us to different b-value regimes in our sample, leading to changing ADC estimates. Fitting our gamma distribution model (Eq. [4]) to the experimental data (b – orange dashed line), we estimate $D_m = 1.50 \cdot 10^{-4}$ mm$^2$/s and $D_s = 2.10 \cdot 10^{-4}$ mm$^2$/s in our postmortem brain sample, with (c) and (d) displaying the resulting diffusivity distribution and displacement profile for these parameters. Error bars in (b) display the standard error of the ADC over the corpus callosum, but are not visible for most flip angles.

*Experimental validation of DW-SSFP flip angle dependency*

Figure 4 reveals the variation in ADC over the corpus callosum (Fig. 4a) of the postmortem sample (blue crosses in Fig. 4b), where the ADC estimated at 90° is almost twice the ADC estimate at 10°, despite no changes in the diffusion encoding of the sequence. This variation in ADC as a function of flip angle is consistent with non-Gaussian diffusion, and inconsistent with Gaussian diffusion, which would correspond to a flat line in Fig 4b. By fitting the ADC estimates to a gamma distribution (dashed orange line), we estimated $D_m = 1.50 \cdot 10^{-4}$ mm²/s and $D_s = 2.10 \cdot 10^{-4}$ mm²/s. The corresponding probability density function and displacement profile are shown in Figs. 4c and d.

**Discussion**

The DW-SSFP signal represents a blurred mixture of signals with well-defined b-values. By defining DW-SSFP derived ADC estimates in terms of an effective b-value, we can transform these estimates into alignment with more conventional diffusion measurements. Monte-Carlo simulations (Fig. 3) yield excellent agreement between simulated signals for a given gamma-distributed system and our forward model. These results suggest that the DW-SSFP signal is able to capture non-Gaussianity and verify the ability to transform DW-SSFP signals into equivalent DW-SE signals. Experimental fitting of our model to data acquired in in the corpus callosum of a whole postmortem brain sample (Fig. 4) demonstrates that use of the original Buxton model produces the predicted flip-angle dependence of the ADC estimate that is expected for non-Gaussian diffusion. Our gamma-distribution model fit (Fig. 4b – dashed orange line) is able to explain this flip-angle dependence of ADC.

The observation of flip-angle-based sensitivity to non-Gaussian diffusion also implies challenges to the use of DW-SSFP. $B_1$-inhomogeneity (e.g. at ultra-high field) translates into varying effective b-value across a sample, leading to spatially varying ADC estimates even when the underlying tissue properties are the same. This confound prevents a simple interpretation of results between, or even within DW-SSFP datasets. One approach is to use the model parameters to derive an ADC map with the same effective b-value within every voxel regardless of local $B_1$, representing a common snapshot of restricted diffusion. This approach could additionally account for the variations in $T_1$, $T_2$ and the diffusivity of tissue, which will also influence the effective b-value (see Supporting Information Fig. S2).

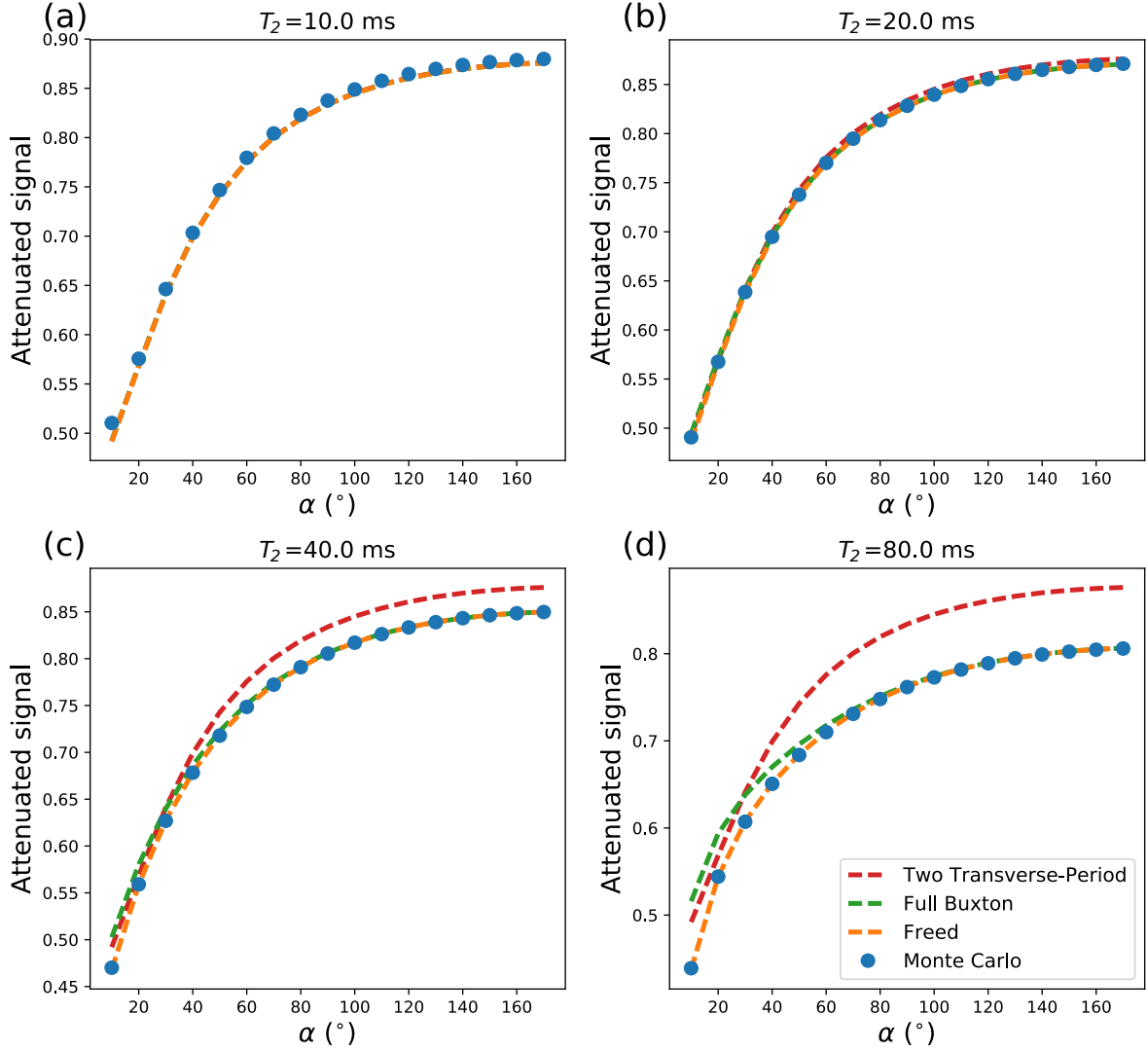

Figure 5: Attenuation of the DW-SSFP signal for four different values of $T_2$, comparing the signal attenuation from Monte-Carlo simulations (blue dots) vs analytical solutions of the two transverse-period (red dashed line) (8), full Buxton (green dashed line) (8) and the Freed (orange dashed line) (18) model assuming a gamma distribution of diffusivities. Simulated parameters (except for $T_2$) are identical to the Monte-Carlo simulations described in the main text. As the $T_2$ estimate increases, we observe a substantial deviation of the signal attenuation predicted by the two transverse-period model vs the Monte-Carlo estimates. Similarly, a deviation is seen with the Full Buxton model, particularly at lower flip angles. As described by Freed (18), under certain experimental regimes the full Buxton model no longer provides accurate estimates of the DW-SSFP signal. The Freed model however provides excellent agreement to the Monte-Carlo simulations across the range of $T_2$ values simulated.

An early version of this framework used the two transverse-period approximation of DW-SSFP (Eq. [1]) to derive analytical solutions (see Appendix) for the ADC and signal under a gamma distribution (30). However, further analysis with Monte-Carlo simulations revealed substantial deviations in signal attenuation when the two transverse-period condition (TR ≥

~1.5·$T_2$ ) is violated (Fig. 5). By utilising numerical integration, we can incorporate other diffusivity distributions without analytical solutions. Moreover, at longer $T_2$, the gamma-distributed Buxton model deviates from Monte-Carlo simulations, whereas the Freed model (18) provided excellent agreement (Fig. 5). In general, the framework presented here is compatible with any DW-SSFP signal model.

One limitation of our study is the lack of comparison between experimental DW-SSFP and DW-SE data. Such a comparison would require acquisition of both DW-SSFP and DW-SE data at multiple flip angles/b-values. However, DW-SE measurements in post-mortem tissue suffer from very low SNR and beyond the scope of this study.

**Conclusion**

By acquiring DW-SSFP data at multiple flip angles, we can probe the non-Gaussian diffusion properties of a sample. We can additionally disentangle the blurred mixture of diffusion weighted signals with different b-values in DW-SSFP. This approach enables the transformation of ADC estimates derived from DW-SSFP to more conventional sequences at a single effective b-value.

**Acknowledgements**

**Appendix**

*Full Buxton model definitions*

$$r = 1 - E_1 \cos\alpha + E_2^2 A_1 A_2^{1/3}(\cos\alpha - E_1),$$

$$s = E_2 A_1 A_2^{-4/3}(1 - E_1 \cos\alpha) + E_2 A_2^{-1/3}(\cos\alpha - E_1).$$

$$F_1 = K - (K^2 - A_2^2)^{1/2}, \qquad [A1]$$

$$K = \frac{1 - E_1 A_1 \cos\alpha - E_2^2 A_1^2 A_2^{-2/3}(E_1 A_1 - \cos\alpha)}{E_2 A_1 A_2^{-4/3}(1 + \cos\alpha)(1 - E_1 A_1)},$$

*Two transverse period approximation of ADC and signal under a gamma distribution*

Under the two transverse period approximation of DW-SSFP (Eq. [1]), we can define:

$$\text{ADC}_{\text{SSFP}} = -\frac{1}{q^2 \text{TR}} \cdot \ln\left[\frac{-(S'_{\text{SSFP}} \cdot E_1 \cos\alpha + 1) + [(S'_{\text{SSFP}} \cdot E_1 \cos\alpha + 1)^2 + 4E_1 \cdot S'_{\text{SSFP}}]^{\frac{1}{2}}}{2E_1}\right], \qquad [A2]$$

where:

$$S'_{\text{SSFP}} = \frac{S_{\text{SSFP}}(\alpha, T_1, T_2, \text{TR}, q, \text{ADC})}{S_{\text{SSFP}}(\alpha, T_1, T_2, \text{TR}, 0, \text{ADC})} \cdot \frac{1 + E_1}{1 - E_1 \cos\alpha}. \qquad [A3]$$

For a gamma distribution:

$$S'_{\text{SSFP},\Gamma} = \left(\frac{D_m}{D_m + q^2 \cdot \text{TR} \cdot D_s^2}\right)^{\frac{D_m^2}{D_s^2}} + E_1 \cdot (1 + \cos\alpha) \cdot \left(\frac{D_m}{q^2 \cdot \text{TR} \cdot D_s^2}\right)^{\frac{D_m^2}{D_s^2}} \cdot \Phi\left(E_1 \cos\alpha, \frac{D_m^2}{D_s^2}, \frac{D_m}{q^2 \cdot \text{TR} \cdot D_s^2} + 2\right), \qquad [A4]$$

where $\Phi$ is the Lerch transcendent (31). Derivations are provided in Supporting Information.


\setcounter{page}{1}
\renewcommand{\thepage}{S\arabic{page}}
\section*{Supporting Information}
\section*{Supporting Figures}
\renewcommand{\thefigure}{S1}
\begin{figure}[H]
\centering
\includegraphics[width = 330pt]{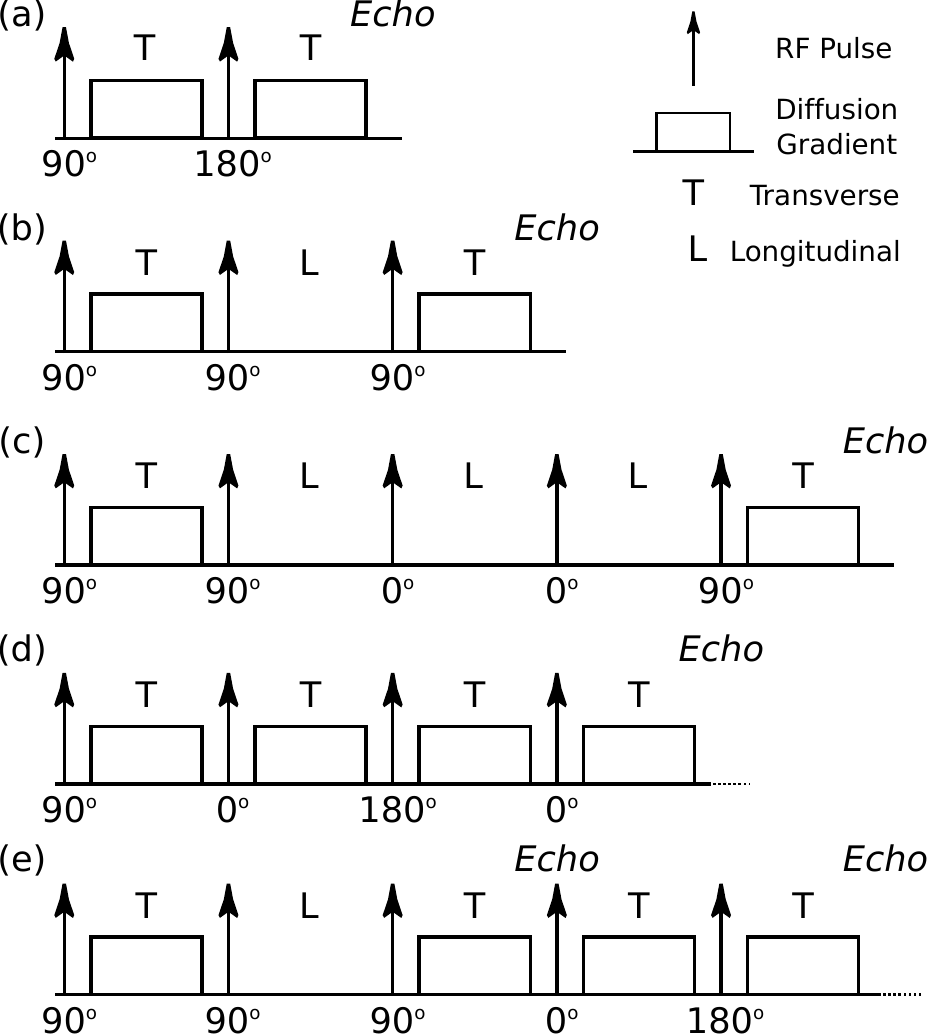}
\caption{In DW-SSFP, repeat application of RF pulses decomposes the the magnetisation into a series of coherence pathways, which are sensitised to the diffusion gradient during transverse-periods. Here we show five example coherence pathways. The spin-echo pathway (a), stimulated-echo pathway (b) and long stimulated-echo pathway (c) only survive for two TRs in the transverse plane, the condition for the two transverse-period approximation (1). These pathways all experience the same q-value, but have different diffusion times, defined as $\Delta = 1\cdot \mathrm{TR}$ (a), $2\cdot \mathrm{TR}$  (b) and $4\cdot \mathrm{TR}$  (c). For the full Buxton model (1) this condition is no longer required, and pathways can experience cumulative sensitisation to the diffusion gradients over multiple TRs, such as the spin-echo pathway in (d), in addition to pathways which generate multiple echoes over their lifetime (e). This leads to pathways with different q-values, in addition to weighting of the signal by $T_{2}$.}
\label{fig:Processing_pipeline}
\end{figure}

\renewcommand{\thefigure}{S2}
\begin{sidewaysfigure}[H]
\centering
\includegraphics[width = 600 pt]{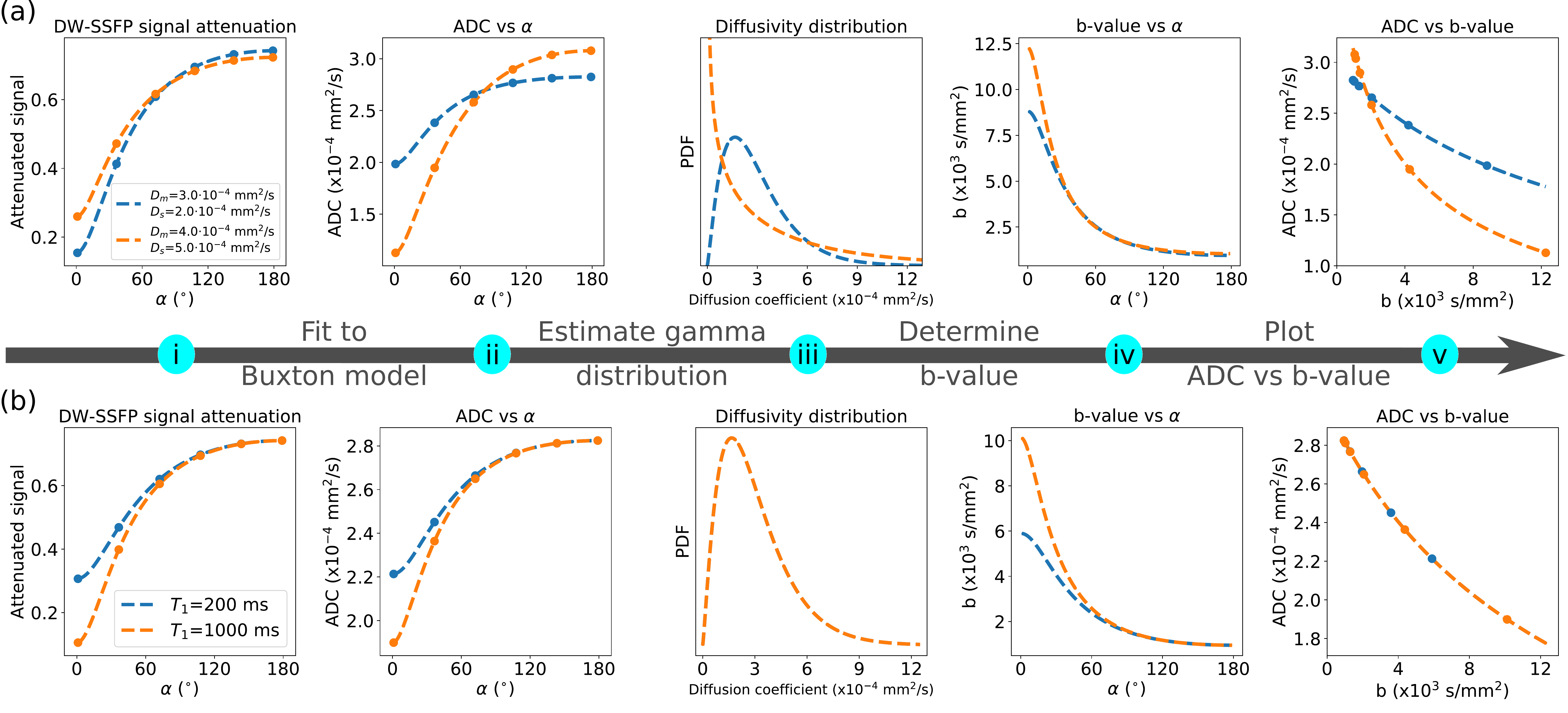}
\caption{Processing pipeline for (a) two samples with different diffusion properties but identical relaxation times and (b) identical diffusion properties but different $T_{1}$ values. Experimental DW-SSFP data is acquired at multiple flip angles (i - dots) and converted into ADC estimates (ii - dots) (Eq. [2] - main text). To avoid fitting for $S_0$, we fit to the DW-SSFP signal attenuation. The DW-SSFP signal model incorporating a gamma distribution of diffusivities (Eq. [4] - main text) is subsequently fit to the ADC estimates at multiple flip angles (by comparing to Eq. [2] - main test) to determine $D_{m}$ and $D_{s}$ (iii). From Eq. [3] in the main text and our fitted values of $D_{m}$ and $D_{s}$, we can simulate the estimated ADC with b-value for a DW-SE sequence by making comparisons with the DW-SE signal under the Stejskal-Tanner model ($S=S_0\exp(-bD)$). From this, we can define an equivalent DW-SE b-value which gives rise to the same ADC estimate at each DW-SSFP flip angle (iv). Our ADC estimates with DW-SSFP can be subsequently plotted vs an effective b-value (v). In (a), this leads to distinct evolution of ADC with effective b-value for the two samples (v). However in (b), the signal evolution is identical (v), despite having a different ADC evolution vs flip angle (ii), reflecting differences in the weighting of the different coherence pathways due to relaxation, leading to different effective b-values along the ADC curve (v - dots).}
\label{fig:pathways}
\end{sidewaysfigure}
\section*{Supporting Tables}
\captionsetup[table]{labelsep=space,justification=raggedright,singlelinecheck=off}
\begin{table}[ht]
\caption*{\textbf{Turbo inversion recovery (TIR) - $T_{1}$ \hspace{18pt} Turbo spin echo (TSE) - $T_{2}$}}
\begin{tabular}{ l l l l}
 Resolution & 0.65 x 0.65 x 1.30 mm$^3$ & \hspace{23pt} Resolution & 0.65 x 0.65 x 1.30 mm$^3$ \\  
 TR & 1000 ms & \hspace{23pt} TR & 1000 ms \\
 TE & 12 ms & \hspace{23pt} TEs & 23, 34, 46, 57, 69 ms \\ 
 TIs & 60, 120, 240, 480, 940 ms & \hspace{23pt} BW & 163 Hz/pixel \\
 BW & 170 Hz/pixel 
\end{tabular}
\end{table}
\renewcommand{\thetable}{S1:}
\begin{table}[ht]
\caption*{\textbf{Actual flip angle imaging (AFI) - $B_{1}$}} 
\begin{tabular}{ l l l }
 Resolution & 1.50 x 1.50 x 1.50 mm$^3$ \\  
 TRs & 7, 21 ms \\
 TE & 2.6 ms \\ 
 BW & 263 Hz/pixel \\
 &
\end{tabular}
\caption{Acquisition protocols for the $T_{1}$, $T_{2}$ and $B_{1}$ maps. Prior to processing, a Gibbs ringing correction was applied to the TIR and TSE data (2). $T_{1}$ and $T_{2}$ maps were derived assuming mono-exponential signal evolution. The $B_{1}$ map was obtained using the methodology described in (3).}
\end{table}
\newpage
\section*{Supporting Derivations}
\section*{The two transverse-period approximation with a gamma distribution of diffusivities}
From Eq. [1] in the main text:
\small
\begin{equation}
    \begin{split}
    S_{\text{SSFP}}&(\alpha,T_{1},T_{2},\text{TR},q,D)= \\ &\frac{-S_{0}(1-E_{1})E_{1}E_{2}^{2}\sin\alpha}{2(1-E_{1}\cos\alpha)}\left[\frac{1-\cos\alpha}{E_{1}}A_{1}+\sin^{2}\alpha\sum_{n=1}^{\infty} (E_{1}\cos\alpha)^{n-1} A_{1}^{n+1}\right],
    \end{split}
    \tag{S1}
    \label{eq:summation_Buxton_2TP}
\end{equation}
\normalsize
\\
where $S_0$ is the equilibrium magnetization, $E_{1}=e^{-\frac{\text{TR}}{T_{1}}}$, $E_{2}=e^{-\frac{\text{TR}}{T_{2}}}$, $\alpha$ is the flip angle, $n$ is the number of TRs between the two transverse-periods for a given stimulated-echo, $A_{\text{1}}=e^{-q^{2}\cdot\text{TR}\cdot{D}}$, $D$ is the diffusion coefficient and $q=\gamma G \tau$, where $\gamma$ is the gyromagnetic ratio, $G$ is the diffusion gradient amplitude and $\tau$ is the diffusion gradient duration. Separating this expression into  spin-echo (SE) and stimulated-echo (STE) pathways:
\small
\begin{equation}
    S_{\text{SE}}=\frac{-S_{0}(1-E_{1})E_{2}^{2}\sin\alpha(1-\cos\alpha)}{2(1-E_{1}\cos\alpha)}\cdot e^{-q^{2}\cdot\text{TR}\cdot{D}},
    \tag{S2}
    \label{eq:summation_Buxton_2TP_SE}
\end{equation}
\normalsize
\\
and:
\small
\begin{equation}
    S_{\text{STE}}=\frac{-S_{0}(1-E_{1})E_{1}E_{2}^{2}\sin\alpha}{2(1-E_{1}\cos\alpha)}\cdot\sin^{2}\alpha \cdot\sum_{n=1}^{\infty} \left[(E_{1}\cos\alpha)^{n-1}\cdot e^{-q^{2}\cdot(n+1)\cdot\text{TR}\cdot D}\right].
    \tag{S3}
    \label{eq:summation_Buxton_2TP_STE}
\end{equation}
\normalsize
\subsubsection{SE term}
Integrating over the SE term with a gamma  distribution of diffusivities:
\small
\begin{equation}
    S_{\text{SE},\Gamma}=\frac{-S_{0}(1-E_{1})E_{2}^{2}\sin\alpha(1-\cos\alpha)}{2(1-E_{1}\cos\alpha)}\cdot\int_{0}^{\infty} e^{-q^{2}\cdot\text{TR}\cdot{D}}\rho(D;D_{m},D_{s})dD,
    \tag{S4}
    \label{eq:SE_integral}
\end{equation}
\normalsize
\\
where $\rho(D;D_{m},D{s})$ is the gamma distribution with mean $D_m$ and standard deviation $D_s$ over $D$. From Eq. [3] in the main text:
\small
\begin{equation}
    \int_{0}^{\infty}e^{-q^{2}\cdot\text{TR}\cdot{D}}\rho(D;D_{m},D_{s})dD=\left[\frac{D_{m}}{D_{m}+q^2\cdot \text{TR}\cdot D_{s}^2}\right]^{\frac{D_{m}^2}{D_{s}^2}}.
    \tag{S5}
    \label{eq:SE_integral_evaluate}
\end{equation}
\normalsize
Therefore:
\small
\begin{equation}
    S_{\text{SE},\Gamma}=\frac{-S_{0}(1-E_{1})E_{2}^{2}\sin\alpha(1-\cos\alpha)}{2(1-E_{1}\cos\alpha)}\cdot\left[\frac{D_{m}}{D_{m}+q^2\cdot \text{TR}\cdot D_{s}^2}\right]^{\frac{D_{m}^2}{D_{s}^2}}.
    \tag{S6}
    \label{eq:SE_gamma_evaluated}
\end{equation}
\normalsize
\\
\subsubsection{STE term}
Integrating over the STE term with a gamma distribution:
\small
\begin{equation}
    S_{\text{STE},\Gamma}=\frac{-S_{0}(1-E_{1})E_{1}E_{2}^{2}\sin\alpha}{2(1-E_{1}\cos\alpha)}\cdot\sin^{2}\alpha \cdot\sum_{n=1}^{\infty} (E_{1}\cos\alpha)^{n-1}\cdot \int_{0}^{\infty}e^{-q^{2}\cdot(n+1)\cdot\text{TR}\cdot D}\rho(D;D_{m},D_{s})dD.
    \tag{S7}
    \label{eq:STE_integral}
\end{equation}
\normalsize
Evaluating the summation term, considering Eq [3] in the main text:
\small
\begin{equation}
    \begin{split}
    & \sum_{n=1}^{\infty} (E_{1}\cos\alpha)^{n-1}\cdot \int_{0}^{\infty}e^{-q^{2}\cdot(n+1)\cdot\text{TR}\cdot D}\rho(D;D_{m},D_{s})dD \\
    = & \sum_{n=1}^{\infty} (E_{1}\cos\alpha)^{n-1}\cdot\left[\frac{D_{m}}{D_{m}+q^2\cdot (n+1) \cdot \text{TR}\cdot D_{s}^2}\right]^{\frac{D_{m}^2}{D_{s}^2}}, \\
    \end{split}
    \tag{S8}
    \label{eq:STE_integral_summation}
\end{equation}
\normalsize
Pulling $D_{m}$ from the numerator and $q^{2}\cdot\text{TR}\cdot D_{s}^2$ from the denominator :
\small
\begin{equation}
    =\left(\frac{D_m}{q^{2}\cdot\text{TR}\cdot D_{s}^2}\right)^{\frac{D_{m}^{2}}{D_{s}^{2}}}\cdot\sum_{n=1}^{\infty} \frac{(E_{1}\cos\alpha)^{n-1}}{\left[\frac{D_{m}}{q^{2}\cdot\text{TR}\cdot D_{s}^2}+(n+1)\right]^{\frac{D_{m}^{2}}{D_{s}^{2}}}}
    \label{eq:STE_integral_summation_pull}
    \tag{S9}
\end{equation}
\normalsize
Rearranging and defining $m=n-1$:
\small
\begin{equation}
    =\left(\frac{D_{m}}{q^{2}\cdot\text{TR} \cdot D_{s}^2}\right)^{\frac{D_{m}^{2}}{D_{s}^{2}}}\cdot\sum_{m=0}^{\infty} \frac{(E_{1}\cos\alpha)^{m}}{\left[(\frac{D_{m}}{q^{2}\cdot\text{TR}\cdot D_{s}^2}+2)+m\right]^{\frac{D_{m}^{2}}{D_{s}^{2}}}}.
    \label{eq:STE_integral_summation_rearrange}
    \tag{S10}
\end{equation}
\normalsize
The summation term is in an equivalent format to the the Lerch Transcendent (4), defined as:
\small
\begin{equation}
    \Phi(z,s,a)=\sum_{m=0}^{\infty} \frac{z^{m}}{\left(a+m\right)^{s}},
    \label{eq:Lerch_trans}
    \tag{S11}
\end{equation}
\normalsize
where $z=E_{1}\cos\alpha$, $s=\frac{D_{m}^{2}}{D_{s}^{2}}$ and $a=\frac{D_{m}}{q^{2}\cdot\text{TR}\cdot D_{s}^2}+2$. Therefore: 
\small
\begin{equation}
    S_{\text{STE},\Gamma}=\frac{-S_{0}(1-E_{1})E_{1}E_{2}^{2}\sin\alpha}{2(1-E_{1}\cos\alpha)}\cdot\sin^{2}\alpha \cdot \left(\frac{D_{m}}{q^{2}\cdot\text{TR}\cdot D_{s}^2}\right)^{\frac{D_{m}^{2}}{D_{s}^{2}}}\cdot\Phi\left(E_{1}\cos\alpha,\frac{D_{m}^{2}}{D_{s}^{2}},\frac{D_{m}}{q^{2}\cdot\text{TR}\cdot D_{s}^2}+2\right).
    \tag{S12}
    \label{eq:STE_integral_evaluated}
\end{equation}
\normalsize
\subsection*{Total signal}
Summing the SE and STE terms:
\small
\begin{equation}
\begin{split}
    S_{\text{SSFP},_{\Gamma}} &(\alpha,T_{1},T_{2},\text{TR},q,D)= \\
    & \frac{-S_{0}(1-E_{1})E_{1}E_{2}^{2}\sin\alpha}{2(1-E_{1}\cos\alpha)}\cdot \left[\frac{1-\cos\alpha}{E_{1}}\cdot\left(\frac{D_{m}}{D_{m}+q^{2}\cdot\text{TR}\cdot D_{s}^2}\right)^{\frac{D_{m}^{2}}{D_{s}^{2}}}+\right. \\
    & \hspace{100pt} \left. \sin^{2}\alpha \cdot \left(\frac{D_{m}}{q^{2}\cdot\text{TR}\cdot D_{s}^2}\right)^{\frac{D_{m}^{2}}{D_{s}^{2}}}\cdot\Phi\left(E_{1}\cos\alpha,\frac{D_{m}^{2}}{D_{s}^{2}},\frac{D_{m}}{q^{2}\cdot\text{TR}\cdot D_{s}^2}+2\right)\right].
    \end{split} 
    \tag{S13}
    \label{eq:Full_2TP_gamma}
\end{equation}
\section*{ADC expression under the two transverse-period approximation}
Summing over Eq. [1] in the main text, noting $\sum_{m=0}^{\infty} r^{m}=\frac{1}{1-r}$, or from (1):
\small
\begin{equation}
    S_{\text{SSFP}}(\alpha,T_{1},T_{2},\text{TR},q,\text{ADC})=-\frac{S_{0}(1-E_{1})(1+E_{1}A_{\text{ADC}})A_{\text{ADC}}(1-\cos\alpha)\sin\alpha}{2(1-E_{1}\cos\alpha)(1-A_{\text{ADC}}E_{1}\cos\alpha)}\cdot E_{2}^{2}.
    \label{eq:Buxton_summed}
    \tag{S14}
\end{equation}
\normalsize
Maintaining terms that depend on ADC:
\small
\begin{equation}
\begin{split}
    \frac{(1+E_{1}A_{\text{ADC}})A_{\text{ADC}}}{(1-A_{\text{ADC}}E_{1}\cos\alpha)} & =  \frac{S_{\text{SSFP}}(\alpha,T_{1},T_{2},\text{TR},q,\text{ADC})}{{S_{\text{SSFP}}(\alpha,T_{1},T_{2},\text{TR},0,\text{ADC})}}\cdot\frac{1+E_{1}}{1-E_{1}\cos{\alpha}} \\
    & = S'_{\text{SSFP}}.\\
    \end{split}
    \label{eq:Full_2TP_gamma_vs_buxton_summed}
    \tag{S15}
\end{equation}
\normalsize
$S_{\text{SSFP}}(\alpha,T_{1},T_{2},\text{TR},q,\text{ADC})$ and $S_{\text{SSFP}}(\alpha,T_{1},T_{2},\text{TR},0,\text{ADC})$ can be substituted by diffusion-weighted and non diffusion-weighted data respectively. 
Multiplying Eq. [\ref{eq:Full_2TP_gamma_vs_buxton_summed}] by the denominator:
\small
\begin{equation}
    (1+E_{1}A_{\text{ADC}})A_{\text{ADC}}-S'_{\text{SSFP}}\cdot(1-A_{\text{ADC}}E_{1}\cos\alpha)=0.
    \label{eq:Full_2TP_gamma_vs_buxton_summed_rearranged}
    \tag{S16}
\end{equation} 
\normalsize
Expanding the brackets and reordering:
\small
\begin{equation}
    E_{1}A_{\text{ADC}}^{2} + (S'_{\text{SSFP}}\cdot E_{1}\cos\alpha+1) A_{\text{ADC}} - S'_{\text{SSFP}} =0.
    \label{eq:Full_2TP_gamma_vs_buxton_summed_rearranged_2}
    \tag{S17}
\end{equation} 
\normalsize
This is a quadratic equation, therefore:
\small
\begin{equation}
    A_{\text{ADC}}=\frac{-(S'_{\text{SSFP}}\cdot E_{1}\cos\alpha+1)\pm[(S'_{\text{SSFP}}\cdot E_{1}\cos\alpha+1)^{2}+4E_{1}\cdot S'_{\text{SSFP}}]^{\frac{1}{2}}}{2E_{1}}.
    \label{eq:Quadratic_pm}
    \tag{S18}
\end{equation}
\normalsize
As $E_{1}$ and $S'_{\text{SSFP}}$ are positive, the numerator is less than $0$ when we consider the negative solution. Considering $A_{\text{ADC}}=e^{-q^{2}\cdot\text{TR}\cdot\text{ADC}}$, this would lead to a complex definition of ADC. Therefore: 
\small
\begin{equation}
    \text{ADC}=-\frac{1}{q^{2}\text{TR}}\cdot\ln{\left[\frac{-(S'_{\text{SSFP}}\cdot E_{1}\cos\alpha+1)+[(S'_{\text{SSFP}}\cdot E_{1}\cos\alpha+1)^{2}+4E_{1}\cdot S'_{\text{SSFP}}]^{\frac{1}{2}}}{2E_{1}}\right]}.
    \label{eq:Quadratic_sol_ADC}
    \tag{S19}
\end{equation}
\normalsize
For a Gamma distribution of diffusivities, by comparing Eq. [\ref{eq:Full_2TP_gamma}] to Eqs. [\ref{eq:Buxton_summed}] and [\ref{eq:Full_2TP_gamma_vs_buxton_summed}]:
\small
\begin{equation}
\begin{split}
    S'_{\text{SSFP},\Gamma}= & \left(\frac{D_{m}}{D_{m}+q^{2}\cdot\text{TR}\cdot D_{s}^2}\right)^{\frac{D_{m}^{2}}{D_{s}^{2}}} + \\
    &  E_{1}\cdot (1+\cos\alpha) \cdot \left(\frac{D_{m}}{q^{2}\cdot\text{TR}\cdot D_{s}^2}\right)^{\frac{D_{m}^{2}}{D_{s}^{2}}}\cdot\Phi\left(E_{1}\cos\alpha,\frac{D_{m}^{2}}{D_{s}^{2}},\frac{D_{m}}{q^{2}\cdot\text{TR}\cdot D_{s}^2}+2\right).
\end{split}
\tag{S20}
    \label{eq:S_primed_gamma}
\end{equation}
\section*{References}
1. Buxton RB. The diffusion sensitivity of fast steady‐state free precession imaging. Magn. Reson. Med. 1993. doi: 10.1002/mrm.1910290212. \\
\\
2. Kellner E, Dhital B, Kiselev VG, Reisert M. Gibbs-ringing artifact removal based on local subvoxel-shifts. Magn. Reson. Med. 2016. doi: 10.1002/mrm.26054. \\
\\
3. Yarnykh VL. Actual flip-angle imaging in the pulsed steady state: A method for rapid three-dimensional mapping of the transmitted radiofrequency field. Magn. Reson. Med. 2007. doi: 10.1002/mrm.21120. \\
\\
4.  Erdilyi A, Magnus W, Oberhettinger F, Tricomi FG. Higher transcendental functions, vol. 1. Bateman Manuscr. Proj. McGraw-Hill, New York 1953:27–31. \\
\end{document}